\documentclass[10pt,journal,epsfig]{IEEEtran}
\usepackage{graphicx}
\usepackage{subfigure,color}
\usepackage{amssymb}
\usepackage{cite,comment}
\usepackage{amsmath}
\usepackage{bm,comment}
\usepackage{algorithm}
\usepackage{algpseudocode}
\makeatletter

\newcommand{\Rmnum}[1]{\expandafter\@slowromancap\romannumeral #1@}
\newcommand{\mv}[1]{\mbox{\boldmath{$ #1 $}}}

\makeatother

\newtheorem{proposition}{Proposition}

\begin{document}
\title{Uplink Cooperative Interference Cancellation for Cellular-Connected UAV: A Quantize-and-Forward Approach}
\author{Weidong Mei and Rui Zhang, \IEEEmembership{Fellow, IEEE}\vspace{-6pt}
\thanks{W. Mei is with the NUS Graduate School for Integrative Sciences and Engineering, National University of Singapore, Singapore 119077, and also with the Department of Electrical and Computer Engineering, National University of Singapore, Singapore 117583 (e-mail: wmei@u.nus.edu).}
\thanks{R. Zhang is with the Department of Electrical and Computer Engineering, National University of Singapore, Singapore 117583 (e-mail: elezhang@nus.edu.sg).}}
\maketitle

\begin{abstract}
Aerial-ground interference is the main obstacle to achieving high spectral efficiency in cellular networks with both traditional terrestrial users and new unmanned aerial vehicle (UAV) users. Due to their strong line-of-sight (LoS) channels with the ground, UAVs could cause/suffer severe interference to/from a large number of non-associated but co-channel base stations (BSs) in their uplink/downlink communications. In this letter, we propose a new cooperative interference cancellation (CIC) scheme for the UAV's uplink communication to mitigate its strong interference to co-channel BSs, which requires only local cooperation between each co-channel BS and its adjacent helping BSs. Specifically, the helping BSs without serving any users in the UAV's communication channel quantize and forward their received signals from the UAV to their aided co-channel BS, which then processes the quantized signals jointly with its own received signal to decode its served terrestrial user's message via canceling the UAV's signal by linear/nonlinear interference cancellation techniques (ICTs). We derive the achievable rates of the proposed CIC scheme with different ICTs as a function of the UAV's transmit power and rate, and thereby unveil the conditions under which the proposed CIC scheme outperforms the existing CIC scheme based on the decode-and-forward (DF) operation of the helping BSs. 
\end{abstract}
\begin{IEEEkeywords}
Cellular-connected UAV, interference cancellation, quantize-and-forward, decode-and-forward.\vspace{-6pt}
\end{IEEEkeywords}

\section{Introduction}
Unmanned aerial vehicle (UAV) (or drone) is an aircraft without requiring any human pilot on board. While UAVs originated from military applications, their use in civilian/commercial applications has been rapidly growing in the last decade, such as for packet delivery, aerial photography, airborne communication, etc\cite{zeng2019accessing}. However, most of today's commercial UAVs are controlled via hand-held radio, which limits their flights to the visual or radio line-of-sight (LoS) range only. To enable beyond visible and radio LoS (BVRLoS) UAV communications, both academia and industry have looked into a new paradigm called \emph{cellular-connected UAV}, by reusing the existing cellular base stations (BSs) and spectrum to serve UAVs and traditional terrestrial user equipments (UEs) at the same time\cite{zeng2019cellular,lin2019mobile,garcia2019essential}. Thanks to the advanced cellular technologies, preliminary field trials\cite{3GPP36777} have demonstrated that the current fourth-generation (4G) long-term evolution (LTE) network is able to meet some basic requirements of UAV-ground communications. However, a more severe aerial-ground interference issue than its counterpart in the existing terrestrial networks also arises\cite{3GPP36777}. Specifically, due to the LoS-dominant channels between UAVs and ground BSs, UAVs may impose/suffer overwhelming interference to/from a large number of BSs in their uplink/downlink communications. As such, effective interference mitigation techniques are required to resolve this new and challenging issue.

\begin{figure}[!t]
\centering
\includegraphics[width=3.0in]{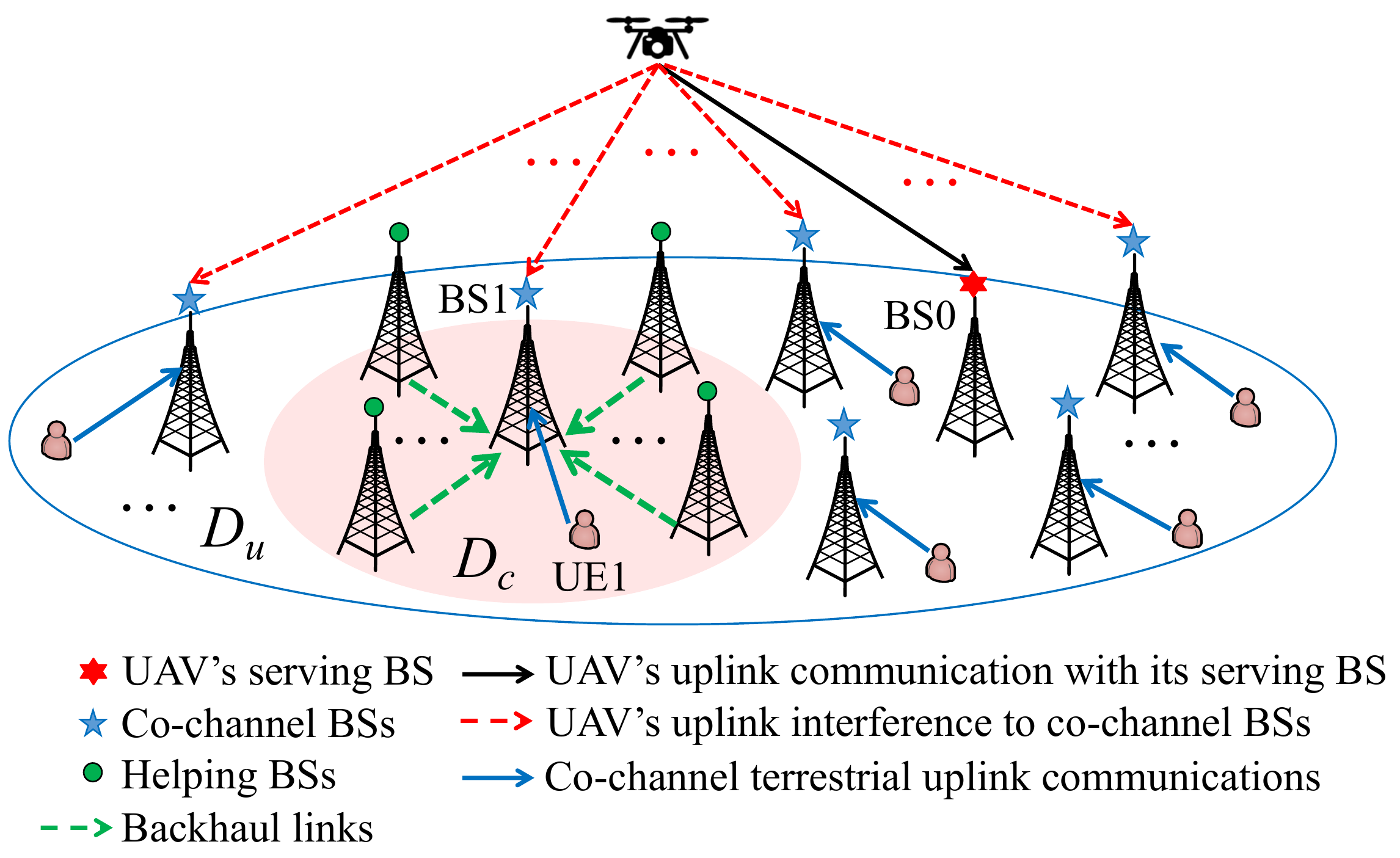}
\DeclareGraphicsExtensions.\vspace{-3pt}
\caption{Uplink UAV communication in a cellular network with cooperative air-to-ground interference cancellation.}\label{UAV_Upcic}
\vspace{-12pt}
\end{figure}
Motivated by the above, this letter aims to investigate the uplink interference mitigation techniques for a cellular network with co-existing UAV and terrestrial UEs, for cancelling the UAV’s interference to a large number of non-associated but co-channel BSs within its wide signal coverage region, denoted by $D_u$ and shown in Fig.\,\ref{UAV_Upcic}. It is worth noting that there have been recent works\cite{cellular2019mei,mei2020uav,liu2019multi,mei2019uplink} on addressing similar problems. In \cite{cellular2019mei} and \cite{mei2020uav}, two new aerial-ground inter-cell interference coordination (ICIC) schemes were proposed for avoiding interference among the UAV and terrestrial UEs by treating the interference as noise. To further improve the rate performance, a decode-and-forward (DF) based cooperative interference cancellation (CIC) scheme was proposed in \cite{liu2019multi} and \cite{mei2019uplink}, by exploiting the existing backhaul links among neighboring BSs. As shown in Fig.\,\ref{UAV_Upcic}, for any co-channel BS (say, BS1) that uses the same communication channel as the UAV for serving its terrestrial UE (say, UE1) and thus suffers the UAV's uplink interference, its adjacent BSs in a local region $D_c$ (termed as {\it helping} BSs) without using this channel (due to fixed/dynamic frequency reuse in existing cellular networks) can decode the UAV's signal and then forward the decoded signal to it for interference cancellation, thus helping improve UE1's achievable rate.

Despite that the above DF-based CIC is effective in general, its performance critically depends on the UAV's transmit power and rate. If the UAV's transmit power is limited as a terminal while its transmit rate is high, which is typical for the high-rate uplink payload communication such as video streaming, none of BS1 and its helping BSs in $D_c$ might be able to decode the UAV's signal, especially when they are far from the UAV (e.g., at the boundary of $D_u$). As a result, UE1 would still suffer severe interference from the UAV at BS1. To deal with the above issue, in this letter we propose a new quantize-and-forward (QF) based CIC scheme. Specifically, the helping BSs in $D_c$ only quantize and forward their received UAV signals to BS1 without decoding them. Then, BS1 can properly combine these quantized signals with its own received signal to either maximize the signal-to-interference-plus-noise ratio (SINR) for decoding the UAV's signal first (by treating UE1's signal as interference) and then cancelling it before decoding UE1's signal (referred to as {\it successive/nonlinear} interference cancellation), or maximize UE1's signal SINR for decoding it directly while linearly suppressing the UAV's interference (refereed to as {\it linear} interference cancellation). With given UAV's transmit power and rate, we derive the optimal combining weights for different interference cancellation techniques (ICTs) and the corresponding UE1's SINRs/achievable rates in closed form. Finally, both analytical and numerical results are provided to unveil the conditions under which the proposed QF-based CIC with different ICTs outperforms the existing DF-based CIC. 

{\it Notations:} Bold symbols in capital letter and small letter denote matrices and vectors, respectively. The conjugate transpose and inverse of a matrix are denoted as ${(\cdot)}^{H}$ and ${(\cdot)}^{-1}$, respectively. $\left[ \cdot \right]_i$ denotes the $i$-th diagonal element of a square matrix. ${\text{diag}}(\mv a)$ denotes a diagonal matrix with its main diagonal given by the vector $\mv a$. ${\mathbb{C}}^n$ denotes the set of complex vectors of length $n$. For a complex number $s$, $\lvert s \rvert$ denotes its norm. ${\mathbb E}[\cdot]$ denotes the expected value of random variables. $x \sim \mathcal{CN}(\mu,\sigma^2)$ means that $x$ is a circularly symmetric complex Gaussian (CSCG) random variable with mean $\mu$ and variance $\sigma^2$. ${\bf 0}_n$ denotes a zero vector of length $n$. 

\section{System Model}
As shown in Fig.\,\ref{UAV_Upcic}, we consider the uplink UAV communication in a subregion $D_u$ of the cellular network, where a UAV UE is associated with a ground BS (named BS0) over a given resource block (RB)\footnote{The proposed scheme can be extended to the case of multiple UAVs if they are assigned with orthogonal RBs to avoid mutual interference.}. Assume that the UAV's transmit power and rate are given as $p_u$ and $r_u$ in its assigned RB, respectively, with $p_u > 0$ and $r_u > 0$, under which its associated BS (BS0) can decode its message reliably. Due to the dense frequency reuse in the cellular network, the UAV may cause strong interference to a large number of non-associated but co-channel BSs over its LoS-dominant channels to them. Without loss of generality, we consider one of such co-channel BSs and its served terrestrial UE (i.e., BS1 and UE1 shown in Fig.\,\ref{UAV_Upcic}), while the results are applicable to other co-channel BSs/UEs similarly.  

Let $h_1$ be the complex-valued baseband equivalent channel from UE1 to BS1, and $f_1$ be that from the UAV to BS1. In general, these channels depend on the BS1/UAV/UE1 antenna gains, path-loss, shadowing and small-scale fading. Note that in the current 4G LTE network, the ground BS antennas are usually tilted downwards for covering the terrestrial UEs and mitigating their inter-cell interference (ICI)\cite{3GPP36777}. As such, we assume that each BS employs an antenna array with fixed directional gain pattern in 3D, while the UAV and terrestrial UEs are assumed to be equipped with a single omnidirectional antenna for simplicity. Let $p_1$ denote the transmit power of UE1. Then, the received signal at its serving BS (BS1) is expressed as
\begin{equation}\label{sum1}
y_1=\sqrt{p_1}h_1x_1+\sqrt{p_u}f_1x_u+z_1,
\end{equation}
where $x_1$ and $x_u$ denote the complex-valued data symbols for UE1 and the UAV with ${\mathbb E}[{\lvert{x_1}\rvert}^2] = 1$ and ${\mathbb E}[{\lvert{x_u}\rvert}^2] = 1$, respectively, and $z_1 \sim \mathcal{CN}(0,\sigma_1^2)$ comprises the background noise and any terrestrial ICI at BS1\footnote{In this letter, we assume that the terrestrial ICI has been well mitigated below the background noise by the existing ICIC techniques in cellular networks, such as cooperative dynamic RB allocation and power control.}, both of which are assumed to be independently Gaussian distributed with $\sigma_1^2$ denoting their total power.

Based on (\ref{sum1}), UE1's SINR at BS1 is given by
\begin{equation}\label{sinr1}
{\gamma}_1 = \frac{p_1\lvert h_1 \rvert^2}{p_u\lvert f_1 \rvert^2+\sigma_1^2}.
\end{equation}
Note that due to the LoS-dominant UAV-BS channel $f_1$, UE1's received signal power, $p_1\lvert h_1 \rvert^2$, may be comparable to or even lower than the UAV's interference power, $p_u\lvert f_1 \rvert^2$, in (\ref{sinr1}). As a result, its SINR/achievable rate may be very low in practice.

\section{Existing DF-Based CIC}
One straightforward method to improve ${\gamma}_1$ is by performing successive/nonlinear interference cancellation (SIC) at BS1 to first decode the UAV's signal, i.e., $\sqrt{p_u}f_1x_u$, and then cancel it in (\ref{sinr1}) for decoding UE1's signal. To this end, the UAV's transmit rate $r_u$ should be no greater than its achievable rate at BS1 by treating UE1's signal as interference/noise, which is given by 
\begin{equation}\label{rate1}
R_1 = {\log_2}\left(1 + \frac{p_u\lvert f_1 \rvert^2}{p_1\lvert h_1 \rvert^2+\sigma_1^2}\right),
\end{equation}
in bits per second per Hertz (bps/Hz). However, due to the comparable power of $p_1\lvert h_1 \rvert^2$ and $p_u\lvert f_1 \rvert^2$, $R_1$ can be lower than $r_u$, making this direct SIC at BS1 infeasible. To overcome this issue, a DF-based CIC scheme was proposed in \cite{liu2019multi} and \cite{mei2019uplink}, where the adjacent helping BSs of BS1 in $D_c$ without serving any UEs in the UAV's assigned RB are exploited to help decoding the UAV's signal. By this means, if one or more helping BSs in $D_c$ are able to decode the UAV's signal and forward it to BS1, BS1 can completely cancel the interference from the UAV before decoding UE1's signal. Note that due to the fixed/dynamic RB allocation in terrestrial ICIC, the adjacent BSs of BS1 usually do not reuse any RB already being used by BS1 to avoid ICI. Hence, they are more likely to be able to decode the UAV's signal as compared to BS1 due to weaker interference from UE1 in general.

In this letter, we assume that BS1 can cooperate with all the BSs in its first $M$-tier neighborhood (i.e., $D_c$), which are given by the set ${\cal J}=\{2,3,4,\cdots,J+1\}$ with $J$ denoting the total number of helping BSs in $\cal J$. Let $f_i$ be the channel from the UAV to BS$i$ with $i \in \cal J$. Then, the received signal at each helping BS$i$ is given by
\begin{equation}\label{sum2}
y_i=\sqrt{p_u}f_ix_u+z_i,
\end{equation}
where $z_i \sim \mathcal{CN}(0,\sigma_i^2)$ comprises the background noise and any terrestrial ICI at BS$i$ (including that from UE1). Thus, the UAV's achievable rate at BS$i$ is given by
\begin{equation}\label{rate2}
R_i = {\log_2}\left(1 + \frac{p_u\lvert f_i \rvert^2}{\sigma_i^2}\right), i \in {\cal J}.
\end{equation}

Define $R_{u,\text{DF}} \triangleq \mathop {\max}\nolimits_{i \in {\cal J} \cup \{1\}} R_i$ as the maximum of the UAV's achievable rates at BS1 and all its helping BSs in $\cal J$. Obviously, the UAV interference can be completely cancelled at BS1 if $R_{u,\text{DF}} \ge r_u$. In this case, the corresponding SINR for decoding UE1's signal at BS1 is expressed as
\begin{equation}\label{sinr2}
{\gamma}_{1,\text{DF}}=
\begin{cases}
\frac{p_1\lvert h_1 \rvert^2}{\sigma_1^2}, &{\text{if}}\;R_{u,\text{DF}} \ge r_u\\
\gamma_1, &\text{otherwise.}
\end{cases}
\end{equation}

By comparing (\ref{sinr2}) with (\ref{sinr1}), it is noted that UE1's SINR can be improved by the DF-based CIC if $R_{u,\text{DF}} \ge r_u$. However, the condition $R_{u,\text{DF}} \ge r_u$ may not hold in practice, e.g., when $r_u$ is high, $p_u$ is low, and/or the UAV is far from BS1 (and thus its helping BSs). In this case, UE1 would still suffer severe SINR loss due to the UAV's uplink transmission. To overcome this limitation, we propose a new QF-based CIC in the next section.\vspace{-6pt}

\section{Proposed QF-Based CIC}
Note that the fundamental limitation of the DF-based CIC lies in the fact that $R_{u,\text{DF}}$ may be limited as compared to $r_u$ due to the individual UAV signal decoding at BS1 or its helping BSs in $\cal J$. In view of this limitation, we propose the following QF-based CIC scheme, where each helping BS$i, i \in \cal J$ only quantizes and forwards $y_i$ to BS1 for joint processing, without the need of individually decoding the UAV's signal.

Mathematically, the quantized signals by all the helping BSs in $\cal J$ are expressed as
\begin{equation}\label{sum3}
\tilde y_i = y_i + e_i = \sqrt{p_u}f_ix_u+z_i+e_i, i \in \cal J,
\end{equation}
where $e_i$ denotes the quantization noise for the received signal at BS$i$ with zero mean and variance $q_i$. We assume that $q_i$’s are independent over $i$ due to independent quantization among the helping BSs. In practice, the quantization noise powers $q_i$'s depend on specific quantization scheme and desired quantization accuracy. If the quantization level is set to be sufficiently high, we have $q_i \approx 0, \forall i \in \cal J$. However, this requires extremely high transmission rate for each helping BS to reliably forward the quantized signals to BS1, which may overwhelm the backhaul links, especially when the UAV UE density is high and the number of helping BSs is large. For example, assume that a simple uniform scalar quantization method is applied by BS$i$ to independently quantize the in-phase and quadrature (I/Q) components of $y_i, i \in \cal J$. Let $D_i$ denote the number of bits that BS$i$ uses to quantize each I/Q component of $y_i, i \in \cal J$. Then, the quantization noise powers for the above quantization method are given by\cite{liu2015joint}
\begin{equation}\label{qerror}
q_i=
\begin{cases}
3(p_u\lvert f_i \rvert^2+\sigma_i^2)2^{-2D_i}, &{\text{if}}\;D_i > 0\\
\infty, &{\text{if}}\;D_i = 0
\end{cases}, i \in \cal J.
\end{equation}
After the quantization, each helping BS$i, i \in \cal J$ forwards $\tilde y_i$ (along with $f_i$, $\sigma_i^2$ and $q_i$) to BS1 for its joint processing via the backhaul link between them. Here, we assume that the transmission rate of each helping BS$i, i \in \cal J$, i.e., $2D_i$ with the above uniform scalar quantization\cite{liu2015joint}, is chosen not to exceed its backhaul capacity. Next, we consider the following nonlinear and linear ICTs (named Scheme 1 and Scheme 2, respectively) for BS1 to cancel the UAV's interference, based on the quantized signals $\tilde y_i, i \in \cal J$ and its received signal $y_1$.\vspace{-6pt}

\subsection{Scheme 1: Nonlinear Interference Cancellation}\label{scheme1}
First, BS1 can properly combine $\tilde y_i, i \in \cal J$ with $y_1$ such that the SINR for decoding the UAV's signal is maximized by treating UE1's signal as noise, so as to facilitate its SIC before decoding UE1's signal. Let ${\mv w}_u \in {\mathbb C}^{J+1}$ be the combining weights applied by BS1. Then, the combined signal is expressed as
\begin{equation}\label{sum4}
{\mv w}^H_u{\mv y} = {\mv w}^H_u{\mv h}x_1 + {\mv w}^H_u{\mv f}x_u+{\mv w}^H_u{\mv z}+{\mv w}^H_u\mv{e},
\end{equation}
where ${\mv y}=[y_1,\tilde y_2,\cdots,\tilde y_{J+1}]^H$, ${\mv h}=[h_1, {\bf 0}_J]^H$, ${\mv f}=[f_1,f_2,\cdots,f_{J+1}]^H$, ${\mv z}=[z_1,z_2,\cdots,z_{J+1}]^H$, and $\mv{e}=[0,e_2,e_3,\cdots,e_{J+1}]^H$.

Based on (\ref{sum4}), the UAV SINR at BS1 is given by
\begin{equation}\label{sinr3}
\gamma_{u,\text{QF1}} = \frac{p_u\lvert {\mv w}^H_u{\mv f}\rvert^2}{{\mv w}^H_u\left(p_1{\mv h}{\mv h}^H+{\text{diag}}({\mv{\Sigma}})+{\text{diag}}(\mv{q})\right){\mv w}_u},
\end{equation}
where ${\mv{\Sigma}}=[\sigma^2_1,\sigma^2_2,\cdots,\sigma^2_{J+1}]$ and $\mv{q}=[0,q_2,\cdots,q_{J+1}]$.

It can be shown that the optimal combining weights that maximize (\ref{sinr3}) are obtained from the well-known minimum mean square error (MMSE) combining\cite{kay1993fundamentals}, i.e.,
\begin{equation}\label{qf.mrc}
{\mv w}_u = (p_1{\mv h}{\mv h}^H+{\mv{\Gamma}})^{-1}{\mv f},
\end{equation}
where ${\mv{\Gamma}} \triangleq {\text{diag}}({\mv{\Sigma}})+{\text{diag}}(\mv{q})$. Correspondingly, the maximum UAV SINR is $\gamma_{u,\text{QF1}}=p_u{\mv f}^H(p_1{\mv h}{\mv h}^H+{\mv{\Gamma}})^{-1}{\mv f}$. Since $p_1{\mv h}{\mv h}^H+{\mv{\Gamma}}$ is a diagonal matrix, it is easy to obtain
\begin{equation}\label{sinr4}
\gamma_{u,\text{QF1}}=\frac{p_u\lvert f_1 \rvert^2}{p_1\lvert h_1 \rvert^2+\sigma_1^2} + \sum\limits_{i \in \cal J}{\frac{p_u\lvert f_i \rvert^2}{\sigma_i^2+q_i}},
\end{equation}
which is the sum of the UAV SINRs in $y_1$ and $\tilde y_i, i \in \cal J$. As a result, with Scheme 1, the UAV's achievable rate at BS1 is given by
\begin{equation}\label{rate3}
R_{u,\text{QF1}} = \log_2\left(1+\frac{p_u\lvert f_1 \rvert^2}{p_1\lvert h_1 \rvert^2+\sigma_1^2} + \sum\limits_{i \in \cal J}{\frac{p_u\lvert f_i \rvert^2}{\sigma_i^2+q_i}}\right).
\end{equation}
and the corresponding SINR for decoding UE1's signal at BS1 is given by
\begin{equation}\label{sinr5}
{\gamma}_{1,\text{QF1}}=
\begin{cases}
\frac{p_1\lvert h_1 \rvert^2}{\sigma_1^2}, &{\text{if}}\;R_{u,\text{QF1}} \ge r_u\\
\gamma_1, &\text{otherwise}
\end{cases}.
\end{equation}

By comparing $R_{u,\text{QF1}}$ in (\ref{rate3}) with $R_{u,\text{DF}}$, it follows that the MMSE combining enables BS1 to jointly decode the UAV's signal based on those received at its helping BSs and itself, instead of decoding it individually under the DF-based CIC. Hence, $R_{u,\text{QF1}}$ and ${\gamma}_{1,\text{QF1}}$ can be improved over $R_{u,\text{DF}}$ and ${\gamma}_{1,\text{DF}}$, respectively, provided that the quantization noise powers $q_i$'s in (\ref{rate3}) are sufficiently small. In particular, if the uniform scalar quantization method is applied, it is easy to verify that $R_{u,\text{QF1}}$ will monotonically increase with $p_u$ and $D_i$'s, by substituting (\ref{qerror}) into (\ref{rate3}). Thus, BS1 is more likely to be able to cancel the interference from the UAV with increasing $p_u$ and $D_i$'s. However, if $D_i$'s are too small, e.g., $D_i=0, \forall i \in \cal J$, we have $R_{u,\text{QF1}}=R_1$, i.e., Scheme 1 reduces to the direct SIC by BS1, i.e., there is no need for QF by the helping BSs.\vspace{-6pt}

\subsection{Scheme 2: Linear Interference Cancellation}\label{scheme2}
Alternatively, BS1 can design ${\mv w}_u$ such that UE1's signal SINR is maximized by linearly suppressing the UAV's interference, so as to decode UE1's signal directly. By this means, Scheme 2 avoids decoding the UAV's signal and thus its performance will not be subjected to the UAV's transmit rate. In addition, it requires a lower-cost linear operation at BS1 as compared to the non-linear SIC in both the DF-based CIC and the QF-based Scheme 1. Based on (\ref{sum4}), the SINR for decoding UE1's signal at BS1 is expressed as
\begin{equation}\label{sinr6}
\gamma_{1,\text{QF2}} = \frac{p_1\lvert {\mv w}^H_u{\mv h}\rvert^2}{{\mv w}^H_u\left(p_u{\mv f}{\mv f}^H+{\mv{\Gamma}}\right){\mv w}_u}.
\end{equation}

Similarly as Scheme 1, the optimal combining weights that maximize (\ref{sinr6}) are MMSE combining, which is given by
\begin{equation}\label{qf.mmse}
{\mv w}_u = (p_u{\mv f}{\mv f}^H+{\mv{\Gamma}})^{-1}{\mv h},
\end{equation}
and correspondingly, the SINR in (\ref{sinr6}) becomes 
\begin{equation}\label{sinr7}
\gamma_{1,\text{QF2}}=p_1{\mv h}^H(p_u{\mv f}{\mv f}^H+{\mv{\Gamma}})^{-1}{\mv h}.
\end{equation}
Unlike Scheme 1, the matrix $p_u{\mv f}{\mv f}^H+{\mv{\Gamma}}$ is not diagonal in (\ref{sinr7}), thus it is generally difficult to express $\gamma_{1,\text{QF2}}$ in a simplified scalar form as in (\ref{sinr4}). However, this can be achieved in our considered scenario due to the special structure of ${\mv h}=[h_1, {\bf 0}_J]^H$, as shown in the following proposition. 

\begin{proposition}
The SINR in (\ref{sinr7}) can be simplified as
\begin{equation}\label{sinr8}
\gamma_{1,\text{QF2}} = \frac{p_1\lvert h_1 \rvert^2}{\sigma_1^2}\left(1-\frac{p_u\lvert f_1 \rvert^2}{\sigma^2_1+p_u\lvert f_1 \rvert^2+p_u\sigma_1^2\sum\limits_{i \in \cal J}{\frac{\lvert f_i \rvert^2}{\sigma_i^2+q_i}}}\right).
\end{equation}
\end{proposition}
\begin{IEEEproof}
It follows from (\ref{sinr7}) that
\begin{align}
\gamma_{1,\text{QF2}} &\mathop = \limits^{(a)} p_1\lvert h_1 \rvert^2[(p_u{\mv f}{\mv f}^H+{\mv{\Gamma}})^{-1}]_{1}\nonumber\\
&\mathop = \limits^{(b)} p_1\lvert h_1 \rvert^2\left[{\mv{\Gamma}}^{-1}-\frac{p_u}{1+p_u{\mv f}^H{\mv{\Gamma}}^{-1}{\mv f}}{\mv{\Gamma}}^{-1}{\mv f}{\mv f}^H{\mv{\Gamma}}^{-1}\right]_1\nonumber\\
&= p_1\lvert h_1 \rvert^2\left(\lbrack \mv{\Gamma}^{-1}\rbrack_1-\frac{p_u\lbrack {\mv{\Gamma}}^{-1}{\mv f}{\mv f}^H{\mv{\Gamma}}^{-1} \rbrack_1}{1+p_u{\mv f}^H{\mv{\Gamma}}^{-1}{\mv f}}\right),\label{eq1}
\end{align}
where equality $(a)$ is due to ${\mv h}=[h_1, {\bf 0}_J]^H$, while equality $(b)$ is derived by invoking the following matrix inverse lemma, i.e., $(\mv A + \mv{BCD})^{-1}=\mv A^{-1}-\mv A^{-1}\mv B(\mv C^{-1}+\mv D\mv A^{-1}\mv B)^{-1}\mv D\mv A^{-1}$ for invertible $\mv A$ and $\mv C$, with setting $\mv A=\mv{\Gamma}$, $\mv B={\mv f}$, $\mv C=p_u$, and $\mv D={\mv f}^H$. Next, it is easy to verify $\lbrack \mv{\Gamma}^{-1}\rbrack_1=\frac{1}{\sigma^2_1}$, $\lbrack {\mv{\Gamma}}^{-1}{\mv f}{\mv f}^H{\mv{\Gamma}}^{-1} \rbrack_1=\frac{\lvert f_1 \rvert^2}{\sigma^4_1}$, and ${\mv f}^H{\mv{\Gamma}}^{-1}{\mv f}=\frac{\lvert f_1 \rvert^2}{\sigma_1^2}+\sum\nolimits_{i \in \cal J}{\frac{\lvert f_i \rvert^2}{\sigma_i^2+q_i}}$. By substituting the above equalities into (\ref{eq1}), we can obtain the SINR expression given in (\ref{sinr8}). The proof is thus completed.
\end{IEEEproof}

It is noted from (\ref{sinr8}) that $\frac{p_1\lvert h_1 \rvert^2}{p_u\lvert f_1 \rvert^2+\sigma_1^2}< \gamma_{1,\text{QF2}}<\frac{p_1\lvert h_1 \rvert^2}{\sigma_1^2}$, i.e., UE1 improves its SINR but still suffers residual interference with the linear interference cancellation. Nonetheless, it can perform better than the QF-based Scheme 1 if $\sum\nolimits_{i \in \cal J}\frac{\lvert f_i \rvert^2}{\sigma_i^2+q_i} \gg \frac{\lvert f_1 \rvert^2}{\sigma_1^2}$, e.g., when the number of helping BSs, $J$, is large enough. However, this comes at the cost of higher cooperation complexity and longer decoding delay. In particular, if the uniform scalar quantization method in (\ref{qerror}) is applied, $\gamma_{1,\text{QF2}}$ can be verified to monotonically increase with $D_i$'s but decrease with $p_u$. In the extreme case of $D_i=0, \forall i \in \cal J$, we have $\gamma_{1,\text{QF2}}={\gamma}_1$ given in (\ref{sinr1}), i.e., there is no need for QF by the helping BSs, similar to Scheme 1.\vspace{-9pt}

\subsection{Performance and Complexity Comparison}\label{comp}
Next, we compare the SINRs/achievable rates for decoding UE1's signal by the considered CIC schemes, namely, the DF-based CIC and the two proposed QF-based CIC schemes (Schemes 1 and 2). The results are summarized as follows.
\begin{itemize}
  \item {\it Scheme 1 versus DF-based CIC}: By comparing (\ref{sinr5}) with (\ref{sinr2}), it follows that the former outperforms the latter if and only if (iff) the UAV's signal can be decoded by the former instead of the latter, i.e., $R_{u,\text{QF1}} \ge r_u>R_{u,\text{DF}}$. Similarly, the latter outperforms the former iff $R_{u,\text{DF}} \ge r_u>R_{u,\text{QF1}}$ (e.g., when the quantization noise powers $q_i, i \in \cal J$ are too high).
  \item {\it Scheme 2 versus DF-based CIC}: Since $\frac{p_1\lvert h_1 \rvert^2}{p_u\lvert f_1 \rvert^2+\sigma_1^2}< \gamma_{1,\text{QF2}}<\frac{p_1\lvert h_1 \rvert^2}{\sigma_1^2}$, the former yields better performance than the latter iff the UAV's signal cannot be decoded by the latter, i.e., $r_u>R_{u,\text{DF}}$. Otherwise, the latter outperforms the former.
  \item {\it Scheme 2 versus Scheme 1}: Similarly as the above, the former outperforms the latter iff the UAV's signal cannot be decoded by the latter, i.e., $r_u>R_{u,\text{QF1}}$. Otherwise, the latter outperforms the former. 
\end{itemize}

In practical implementation, Scheme 2 has the lowest decoding complexity among the three considered schemes due to the linear ICT applied at BS1. However, both Schemes 1 and 2 yield higher signaling overhead than the DF-based CIC, since the quantized signals by all helping BSs need to be forwarded to BS1; while in the latter case, only the decoded UAV signal at one of the helping BSs needs to be forwarded.\vspace{-6pt}
\begin{figure*}[hbtp]
\centering
\subfigure[]{\includegraphics[width=0.32\textwidth]{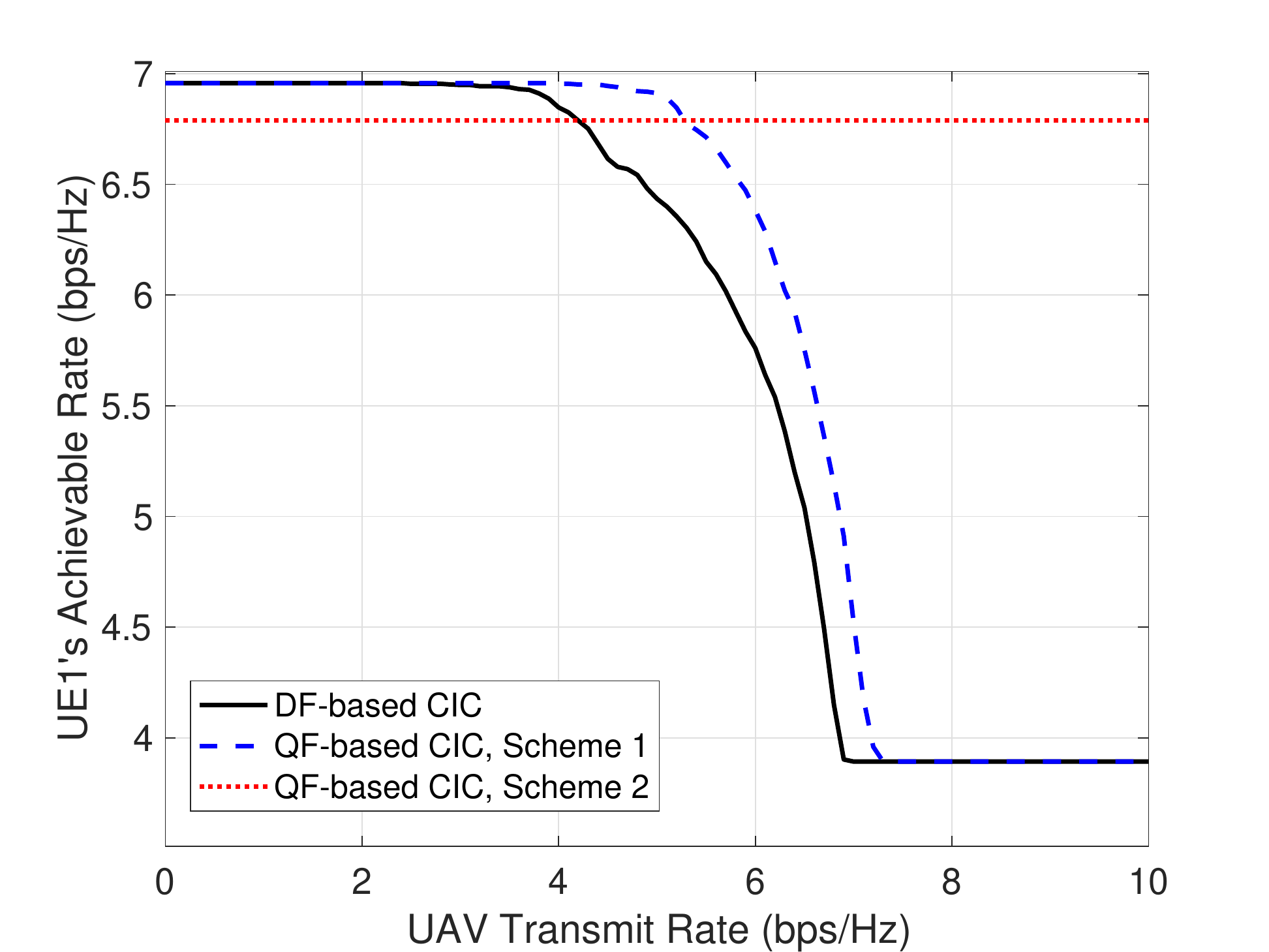}}
\subfigure[]{\includegraphics[width=0.32\textwidth]{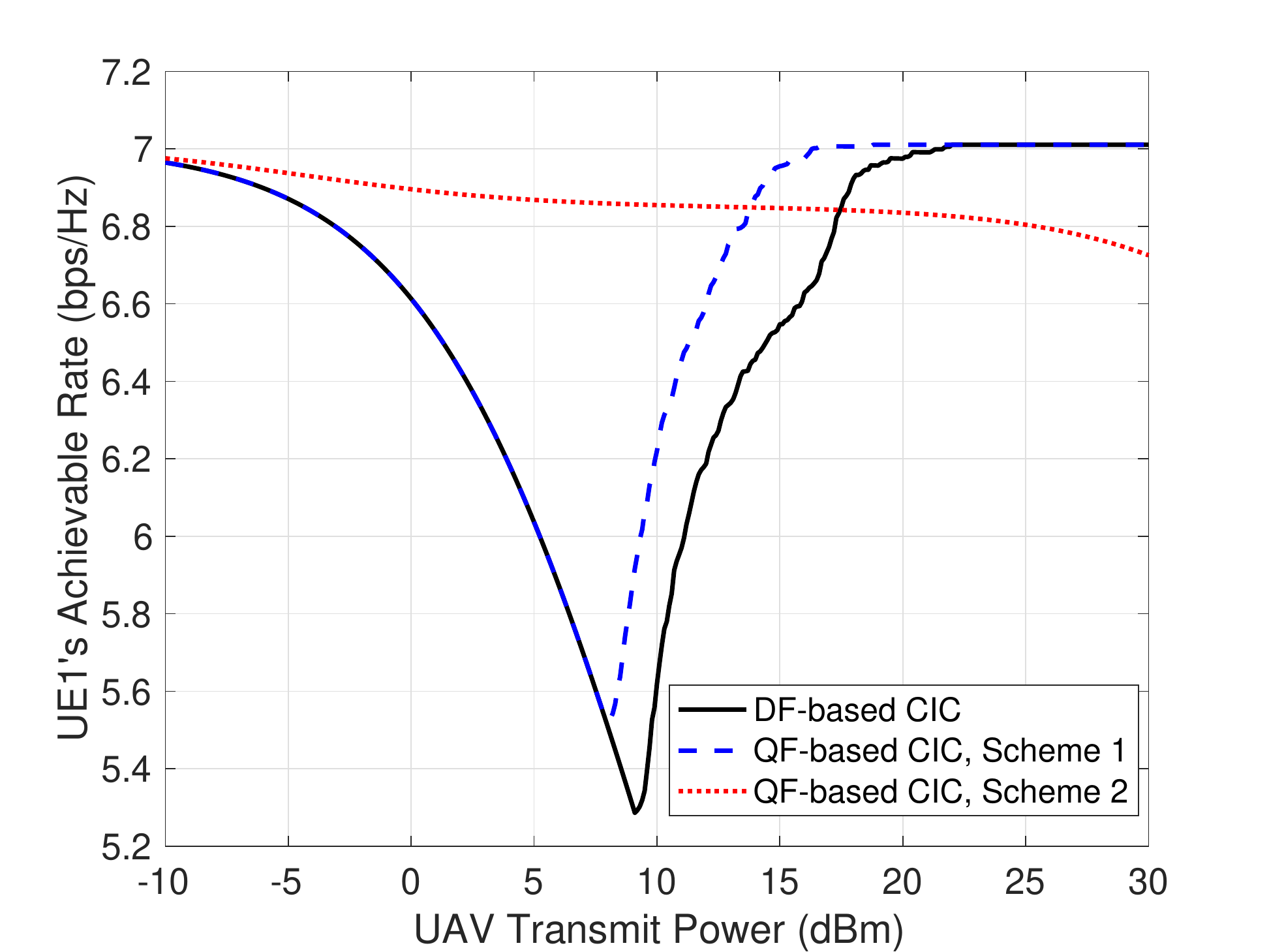}}
\subfigure[]{\includegraphics[width=0.32\textwidth]{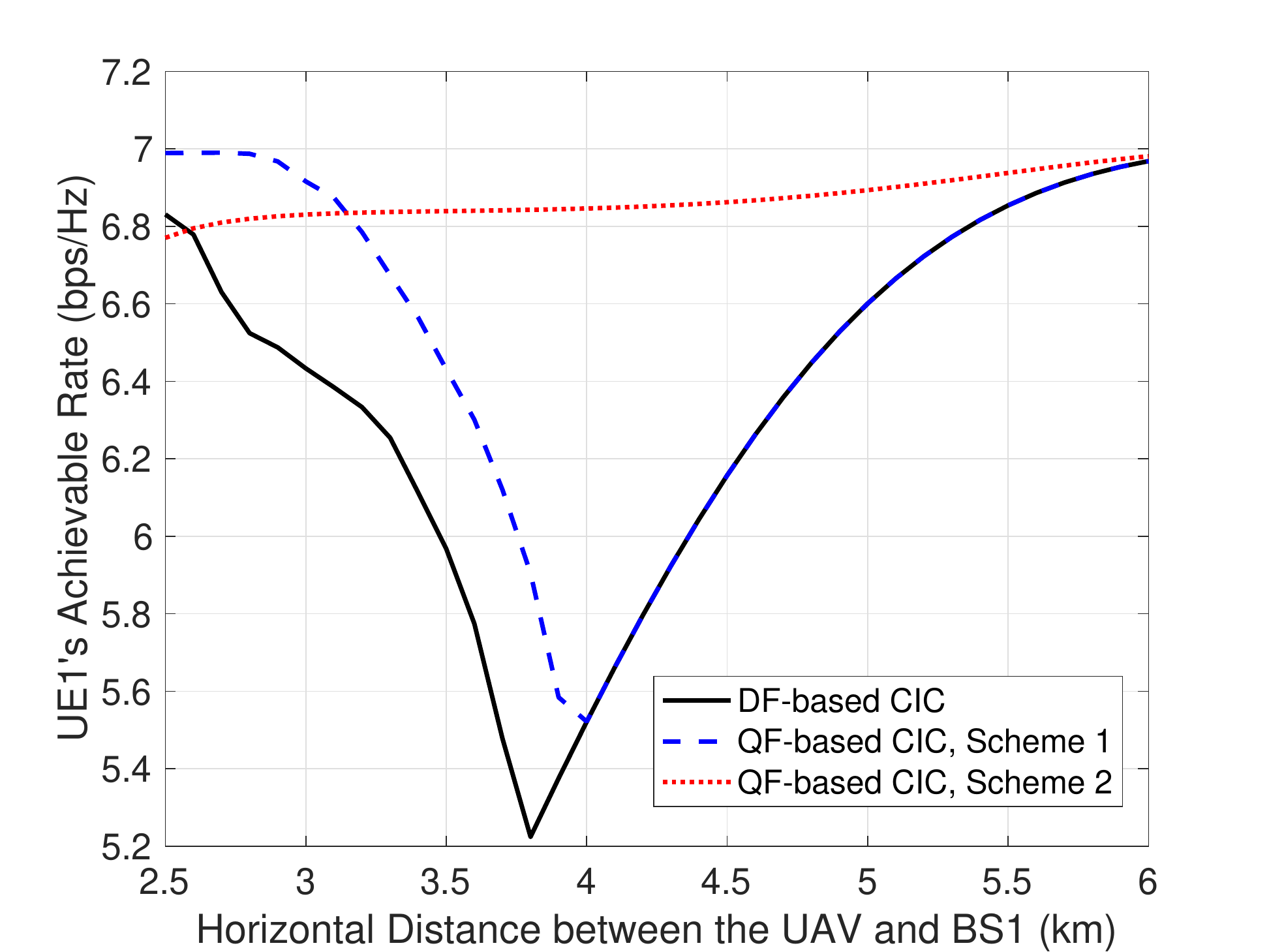}}\vspace{-3pt}
\caption{UE1's achievable rate versus (a) the UAV's transmit rate; (b) the UAV's transmit power; and (c) the UAV's horizontal distance with BS1.}\label{sim}\vspace{-12pt}
\end{figure*}

\section{Numerical Results}
In this section, numerical results are provided to evaluate the performance of the proposed QF-based CIC (Schemes 1 and 2) as compared to the existing DF-based CIC. Unless otherwise specified, the simulation settings are as follows. The shape of each cell is assumed to be hexagonal. The helping BSs of BS1 constitute its nearest adjacent BSs in all considered CIC schemes, i.e., $M=1$ and $J=6$. The bandwidth of the UAV's assigned RB is set to 180 kHz. The carrier frequency $f_c$ is $2$ GHz, and the noise power spectrum density at BS receiver is $-164$ dBm/Hz. For the terrestrial channels, their path-loss and shadowing are modeled based on the urban macro (UMa) scenario in the 3GPP report\cite{3GPP36777}, while the small-scale fading is modeled as Rayleigh fading. The cell radius is set to 800 in meter (m), and the heights of all BSs are 25 m. The altitude of the UAV is fixed as 200 m, while its horizontal distance with BS1 is 3 km. The BS antenna pattern is assumed to be directional in the vertical plane with 10-degree downtilt angle. The total number of non-associated but co-channel BSs in $D_u$ is 8 (including BS1), and their served terrestrial UEs' locations are randomly generated in their located cells. All BS-UAV channels follow the probabilistic LoS channel model in the UMa scenario in \cite{3GPP36777}. The transmit power for all terrestrial UEs is assumed to be identical as $p_1=15$ dBm. We consider that all helping BSs adopt 6-bit uniform scalar quantization to quantize their received UAV signals, i.e., $D_i=6, \forall i \in \cal J$ in (\ref{qerror}). All the results shown below are averaged over 1000 random channel and location realizations of terrestrial UEs.

First, Fig.\,\ref{sim}(a) shows UE1's achievable rate (defined as $\log_2(1+\text{SINR})$ in bps/Hz, where SINR denotes the UE1's achievable SINR in each scheme) versus the UAV's transmit rate, $r_u$. The UAV's transmit power is 15 dBm. It is observed that UE1 can attain its maximum achievable rate in both the DF-based CIC and the QF-based CIC Scheme 1 when $r_u$ is low. This is because the first conditions in (\ref{sinr2}) and (\ref{sinr5}) are both met in this case. Thus, BS1 can completely remove the interference from the UAV. However, as $r_u$ increases, these conditions no longer hold in more channel realizations; thus, UE1's achievable rates by both schemes are observed to decrease and finally attain the minimum given by $\log_2(1+\gamma_1)$. Nonetheless, it is observed that the QF-based CIC Scheme 1 yields a better performance than the DF-based CIC when $r_u$ is sufficiently high, i.e., it is more unlikely for the helping BSs to individually decode the UAV's signal. In contrast, without the need of decoding the UAV's signal, UE1's achievable rate is observed to be unaffected by $r_u$ in the QF-based CIC Scheme 2. It is also observed that this scheme significantly outperforms the other two schemes when $r_u$ is sufficiently high, in accordance with the conditions given in Section \ref{comp}. 

Next, in Fig.\,\ref{sim}(b), we plot UE1's achievable rate versus the UAV's transmit power, $p_u$, with fixed $r_u=5$ bps/Hz. It is observed that with increasing $p_u$, UE1's achievable rate by Scheme 2 decreases, which is consistent with our discussion at the end of Section \ref{scheme2}. Whereas for the other two schemes, UE1's achievable rates are observed to first decrease and then increase with $p_u$. This is because when $p_u$ is small, the first conditions in (\ref{sinr2}) and (\ref{sinr5}) cannot be met. As a result, UE1's achievable rates are degraded by increasing $p_u$ due to higher interference power. However, as $p_u$ further increases, the above conditions are more likely to be met, and thus UE1's achievable rates start to increase since a larger $p_u$ facilitates decoding the UAV's signal in both schemes. It is also observed that the proposed QF-based CIC Schemes 1 and 2 outperform the DF-based CIC when $p_u$ is high and low, respectively, which are consistent with the conditions given in Section \ref{comp}. 

Last, we plot UE1's achievable rate versus the UAV's horizontal distance with BS1 in Fig.\,\ref{sim}(c), with $r_u=5$ bps/Hz and $p_u=15$ dBm. It is observed that UE1's achievable rate by Scheme 2 increases when the UAV is further away from BS1. This is expected as the UAV's interference to UE1 at BS1 is weakened due to its increased distance with BS1. However, UE1's achievable rates by the other two schemes are observed to first decrease and then increase. This is because increasing the distance between the UAV and BS1 (and its helping BS$i, i \in \cal J$) also degrades BS1's interference cancellation capability (thus, $R_{u,\text{DF}}$ and $R_{u,\text{QF1}}$) due to the reduced channel power gains, $\lvert f_i \rvert^2, i \in \cal J$. It is observed that when the UAV-BS1 horizontal distance is smaller than 3.8 km (4 km) in Scheme 2 (DF-based CIC), BS1's degraded interference cancellation capability dominates over the reduced UAV interference at BS1. However, if this distance continues to increase, the latter will become dominant, thus resulting in the improvement of UE1's achievable rates. \vspace{-6pt}

\section{Conclusions}
This letter proposes a new QF-based CIC scheme with different ICTs to mitigate the strong uplink interference from the UAV to a large number of co-channel BSs serving terrestrial UEs, by exploiting the quantized signals shared by their adjacent helping BSs. Analytical and numerical results both reveal that the proposed QF-based scheme can yield significant performance gains over the existing DF-based CIC scheme for practical UAV's transmit rate/power as well as channels with BSs, while the use of linear or nonlinear ICT for the QF-based CIC should be decided based on the actual condition.\vspace{-3pt}

\bibliography{UAV_QF}
\bibliographystyle{IEEEtran}
\end{document}